%% file: main.tex
\documentclass{Interspeech}
\usepackage{multirow}
\usepackage{subcaption}
\usepackage{caption}
\usepackage[
   backend=biber,
   style=ieee,
   citestyle=numeric-comp,
   maxbibnames=10, minbibnames=10,
   maxcitenames=3,
   doi=false,isbn=false,url=false,eprint=false
]{biblatex}
\input{sourcemap_definitions.tex}
\defbibheading{bibliography}[\refname]{}

\interspeechcameraready

\title{HASRD: Hierarchical Acoustic and Semantic Representation Disentanglement}

\author[affiliation={1,2}]{Amir}{Hussein}
\author[affiliation={1}]{Sameer}{Khurana} 
\author[affiliation={1}]{Gordon}{Wichern} 
\author[affiliation={1}]{Francois G.}{Germain}
\author[affiliation={1}]{Jonathan}{Le Roux}

\affiliation{}{Mitsubishi Electric Research Laboratories}{USA}
\affiliation{}{Johns Hopkins University}{USA}
\email{ahussei6@jhu.edu, \{wichern,germain,leroux\}@merl.com}


\begin{document}

\maketitle

\begin{abstract}
  Effective speech representations for spoken language models must balance semantic relevance with acoustic fidelity for high-quality reconstruction. However, existing approaches struggle to achieve both simultaneously. To address this, we introduce Hierarchical Acoustic and Semantic Representation Disentanglement (HASRD, pronounced ``hazard''), a framework that factorizes self-supervised learning representations into discrete semantic and acoustic tokens. HASRD assigns the semantic representation to the first codebook, while encoding acoustic residuals in subsequent codebooks. This preserves ASR performance while achieving high-quality reconstruction. Additionally, we enhance HASRD’s encoder efficiency, improving ASR performance without compromising reconstruction quality. Compared to SpeechTokenizer, HASRD achieves a $44$\% relative WER improvement, superior reconstruction quality, and $2$x lower bitrate, demonstrating its effectiveness in disentangling acoustic and semantic information. 
\end{abstract}

\begingroup
\makeatletter
\renewcommand{\footnoterule}{}
\renewcommand\@makefnmark{}
\renewcommand{\@makefntext}[1]{\noindent#1}
\makeatother
\footnotetext[0]{Work done while A. Hussein was an intern at MERL.}
\renewcommand\thefootnote{}\footnote{Submitted to Interspeech 2025.}
\endgroup

\section{Introduction}
The remarkable performance of large language models has inspired the development of spoken language models (SLMs) that leverage discrete speech representations \cite{ xu2024comparing, borsos2023audiolm, peng2024survey}. SLM use cases range from cross-modal conversational abilities, enabling them to understand and generate multimodal content, to various speech-related tasks such as automatic speech recognition (ASR), speech translation (ST), and spoken language understanding \cite{zhang2023speechgpt, peng2024survey}.
Discrete speech representations for SLMs are broadly categorized into semantic and acoustic tokens \cite{zhang2023speechtokenizer,borsos2023audiolm}. Prior work \cite{zhang2023speechtokenizer, chang2024exploring} highlights the importance of balancing semantic content and acoustic fidelity to support both language modeling and high-quality synthesis. Semantic tokens are derived by pre-training a transformer encoder via self-supervised learning (SSL) typically based on masked prediction \cite{hsu2021hubert, chen2022wavlm, chung2021w2v}, followed by vector quantization. While effective for ST and ASR \cite{chang2023exploration, chang2024exploring, polyak21_interspeech}, semantic tokens often degrade synthesis quality and lack fine acoustic details compared to neural audio codecs \cite{zhang2023speechtokenizer, nguyen2023expresso}. In contrast, acoustic tokens are produced by neural audio codecs trained for reconstruction \cite{zeghidour2021soundstream, defossez2022high}, which use hierarchical quantization such as residual vector quantization (RVQ) to preserve audio fidelity \cite{defossez2024moshi, kumar2023high}, but lack semantic structure. This trade-off motivates the need for a unified architecture that supports both accurate recognition and high-quality resynthesis. 

Recently, researchers proposed jointly learning semantic and acoustic discrete representations by distilling semantic information into the first codebook of a neural codec, while using the remaining codebooks for acoustic reconstruction \cite{ye2024codec,zhang2023speechtokenizer,defossez2024moshi}. While this improves semantic content, performance on semantic tasks still degrades compared to the teacher model. Moreover, state-of-the-art codecs \cite{kumar2023high} achieve high reconstruction quality by using codebooks with small latent dimensionalities, optimizing codebook utilization. However, this design makes distillation within such constrained latent spaces particularly challenging. A closely related work \cite{sim2024skqvc}, developed concurrently with ours, focuses on voice conversion by disentangling content (K-means quantized SSL) from speaker variation (SSL residual after content subtraction). However, it assumes that a single SSL layer captures both content and acoustic information, and it remains unclear how well the learned representation performs on semantic tasks such as ASR. 

To address the aforementioned limitations, we propose a unified model that disentangles and jointly learns semantic and acoustic speech tokens. Unlike prior work, our approach aims to preserve performance on semantic tasks while achieving high-quality reconstruction. To this end, we introduce a novel hierarchical disentanglement framework inspired by the concept of RVQ. Prior research has demonstrated that SSL representations are well-suited for semantic tasks and can also facilitate speech resynthesis \cite{shi24h_interspeech, yang2024towards, hsu2023revise, mu2024self}. Motivated by this, we factorize SSL representations into discrete semantic and acoustic tokens, where the semantic layer forms the first RVQ codebook, and the remaining codebooks encode the acoustic residual via a reconstruction objective.
Our key contributions are: 1) HASRD, a novel framework for learning disentangled acoustic and semantic representations, 2) an efficient CNN encoder that reduces computational complexity while maintaining high reconstruction quality in neural codecs, and 3) an improved design of SSL that utilizes random projection quantization, leading to better quantized representations. Additionally, we provide an in-depth analysis of information disentanglement, highlighting the conflicting objectives of acoustic and semantic representations and explaining how optimizing one often degrades the other.
\section{Proposed Approach}
\begin{figure*}
  \centering
  \begin{subfigure}[b]{0.4\textwidth}
  \centering
  \includegraphics[width=7cm,height=2.5cm]{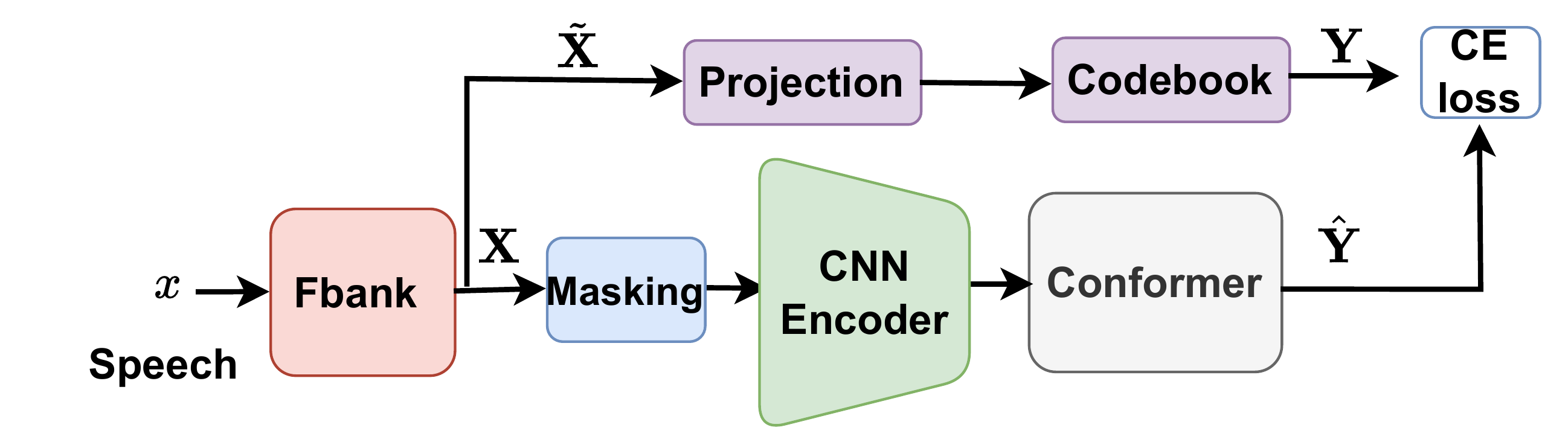}
  \vspace{-5mm}
  \caption{BestRQ pretraining}
  \vspace{-2mm}
  \label{fig:BestRQ-pretraining}
  \end{subfigure}
  \hspace{5pt}
  \begin{subfigure}[b]{0.55\textwidth}
  \centering
  \includegraphics[width=10cm,height=2.8cm]{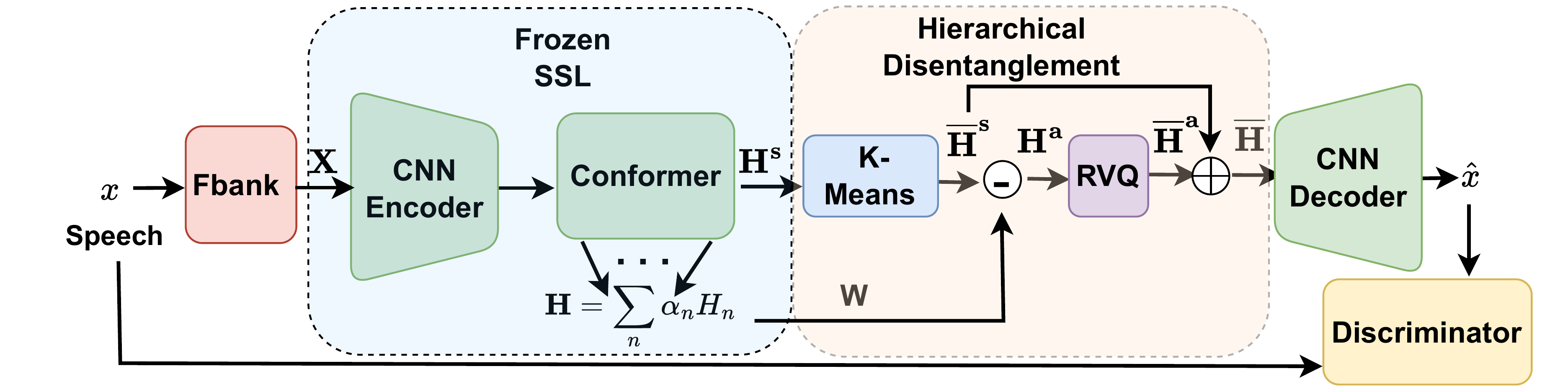}
  \vspace{-5mm}
  \label{fig:continued-training}
  \caption{Continued training with reconstruction objective}
  \vspace{-2mm}
  \end{subfigure}
  \caption{Proposed HASRD framework for hierarchical disentangling of acoustic and semantic representations}
  \label{fig:proposed-method}
  \vspace{-0.4cm}
\end{figure*}
For discrete speech representations to be effective for SLMs, they must capture semantic information while preserving acoustic details to enable high-quality speech synthesis. However, these objectives often conflict, as enhancing one can degrade the other, making it challenging to learn a single representation that balances both, as illustrated in Section~\ref{sec:Disentanglement}. To overcome this challenge, we propose a two-stage approach that combines 1) SSL pretraining with a masked language modeling (MLM) objective and 2) continued training with residual vector quantization (RVQ) and reconstruction objective. This framework enables good word error rate (WER) performance, strong compression, and high-quality reconstruction.

\subsection{Semantic Pre-Training}\label{sec:ssl}
We adopt SSL pre-training with an MLM objective. Our CNN encoder design is inspired by the Descript Audio Codec framework \cite{kumar2023high}. Given computational constraints, we first enhance the efficiency of the CNN encoder. Instead of raw audio, our system first computes an $F$-dimensional log mel-spectrogram $\mathbf{X} \in \mathbb{R}^{T \times F}$ as input. The encoder consists of two 1D-CNN blocks with strided convolutions, each containing three residual units with 1D dilated convolutions. We use the Snake activation function \cite{kumar2023high} for nonlinearity. To further reduce the number of parameters, we employ depthwise separable convolutions \cite{liu2022convnet}, a special case of grouped convolution where the number of groups matches the number of channels. To better capture semantic information, we enhance the CNN encoder by stacking Conformer modules with multi-head attention (MHA) on top of the CNN blocks, and we pre-train it with an MLM objective. Following BestRQ~\cite{chiu2022self}, we apply a random projection quantizer to the stacked fbank feature sequence $\tilde{X} \in \mathbb{R}^{T^{\prime}\times F^{\prime}}$, where $T^{\prime} = T/4$ and $F^{\prime} = 4F$, with a single codebook to generate labels $\mathbf{Y}$ for masked speech features. The stacked features at each frame
are quantized to the nearest codebook entry using the standard cosine-distance-based code lookup. Note that the time resolution $T^{\prime}$ matches the output of the CNN encoder in our SSL model.
The SSL model is trained to predict labels for the masked regions using unmasked features as context, guided by quantizer-provided labels:
{\setlength{\abovedisplayskip}{3pt}%
 \setlength{\belowdisplayskip}{3pt}%
 \begin{equation}
\mathcal{L}_{\text{MLM}} = - \sum_{m \in \mathcal{M}} \log P(y_{m} \mid \mathbf{X}_{\setminus \mathcal{M}}),
\end{equation}}
where $\mathcal{M}$ denotes the set of masked positions and 
$\mathbf{X}_{\setminus \mathcal{M}}$ the unmasked features.

\subsection{Hierarchical Information Disentanglement}\label{sec:factorization}
It has been shown that the representations of specific SSL layers are well-suited for semantic tasks, while others are more effective for acoustic-related tasks such as speech resynthesis \cite{shi24h_interspeech, polyak21_interspeech, yang2024towards}. Motivated by this observation, we hypothesize that SSL representations can be factorized into semantic and acoustic representations. Let $\{\mathbf{H}_1, \dots, \mathbf{H}_N\}$, where $\mathbf{H}_n \in \mathbb{R}^{U \times D}$, denote the latent representations with temporal dimension $U$ (after CNN downsampling) and latent dimension $D$, obtained from the pre-trained SSL layers described in Section~\ref{sec:ssl}. Since it is not known which representation is best suited for learning acoustic tokens, we propose to introduce learnable weights that adaptively combine the representations as shown in Fig.~\ref{fig:proposed-method}.
Specifically, we compute the full representation $\mathbf{H}$ as a convex combination $\mathbf{H} = \sum_{n=1}^{N} \alpha_n \mathbf{H}_n$, where $\alpha_n$ is a learned normalized scalar weight. We identify the index $n_s \in \{1, \dots, N\}$ of the best performing representation for the ASR task, and define the corresponding representation $\mathbf{H}^s=\mathbf{H}_{n_s}$ 
as the semantic representation.
The corresponding quantized semantic representation is then obtained as $\overline{\mathbf{H}}^{s} = \operatorname{kmeans}(\mathbf{H}^s)$, where $k$-means is trained offline as described in Section~\ref{sec:setup}.
The acoustic representation $\mathbf{H}^a \in \mathbb{R}^{U \times D}$ is obtained by removing the semantic component from a projected version of the full representation $\mathbf{H}$ using a learnable transformation matrix $\mathbf{W} \in \mathbb{R}^{D \times D}$, intended to align the semantic part of $\mathbf{H}$ with $\overline{\mathbf{H}}^{s}$ for optimal removal:
\begin{equation}
\mathbf{H}^a = \mathbf{H}\mathbf{W} - \overline{\mathbf{H}}^{s}
\label{eq:1}
\end{equation}
During the continued training stage, the encoder (CNN + Conformer) is frozen. To ensure that the decoder reconstruction relies on the semantic quantized representation $\overline{\mathbf{H}}^{s}$, we adopt a multistage hierarchical quantization approach inspired by RVQ. Specifically, we treat $\overline{\mathbf{H}}^{s}$ as the output of the first quantization codebook of an RVQ, and quantize the residual $\mathbf{H}^{a}$ obtained by Eq.~(\ref{eq:1}) using subsequent RVQ codebooks, denoting by ${\overline{\mathbf{H}}^{a}_m}$ the output of each quantization stage $m$, $m=1,\dots,M-1$. The final quantized representation is then computed as:
{\setlength{\abovedisplayskip}{3pt}%
 \setlength{\belowdisplayskip}{3pt}%
 \begin{equation}
\overline{\mathbf{H}} = \overline{\mathbf{H}}^{s} + \sum_{m=1}^{M-1} \overline{\mathbf{H}}^{a}_m.
\label{eq:2}
\end{equation}}
The resulting $\overline{\mathbf{H}}$ is passed to the decoder, which mirrors the encoder's design but employs 1D transposed convolutions to progressively upsample the feature sequence into a waveform.

\vspace{-.05cm}
\subsection{Discriminators and Reconstruction Loss Functions}
Our approach adopts the discriminators and loss functions proposed in \cite{kumar2023high}. Specifically, we employ a combination of multiscale (MSD) \cite{kumar2019melgan} and multiperiod (MPD) waveform discriminators \cite{kong2020hifi}, alongside a complex short-time Fourier transform (STFT) discriminator operating at multiple timescales. To further enhance perceptual quality, we integrate a multiband, multiscale STFT discriminator with an L1 feature matching loss \cite{kumar2023high}. 
For reconstruction, we optimize an L1 loss on mel spectrograms, leveraging multiple window lengths and mel-bin configurations to capture both fine and coarse spectral details. Codebook learning combines commitment and codebook losses, with gradient propagation handled via the straight-through estimator.

\begin{table*}[tb!]
\centering
\caption{Impact of the improved CNN encoder on reconstruction quality (mel distance, STFT distance, and perceptual quality via VISQOL) and computational complexity (reduction of GPU memory usage and improvement of inference speed)}\vspace{-8pt}
\resizebox{\textwidth}{!}
{%
\begin{tabular}{lccccccccc}
\toprule
\textbf{Model} & {\textbf{Segment}} & \textbf{Bitrate} & \textbf{Train} & \textbf{Params} & \textbf{\!GPU-Mem\!} & \textbf{\!Inference\!} & \textbf{Mel $\downarrow$} & \textbf{STFT $\downarrow$} & \textbf{ViSQOL $\uparrow$} \\
\midrule
DAC                  & 0.4 s & 2.79 kbps & 100 h & 76.6M  & -      & -  & 0.75 & 1.55 & 4.22 \\
DAC (baseline)                & 5.0 s  & 2.79 kbps & 100 h  & 76.6M  & -      & -  & \textbf{0.66} & \textbf{1.45} & 4.45 \\
DAC+fbank            & 5.0 s  & 2.79 kbps & 100 h & 68.0M  & $-$25\%  & x2 & \textbf{0.66} & 1.50 & \textbf{4.50} \\
DAC+fbank+SepConv    & 5.0 s  & 2.79 kbps & 100 h & 48.0M  & $-$25\%  & x3 & 0.68 & 1.52 & 4.45 \\

\bottomrule
\end{tabular}%
}
  \vspace{-0.3cm}
\label{tab:efficient}
\end{table*}

\section{Experimental Setup}
\label{sec:setup}
\textbf{Data \& Pre-processing:}  
We train our models on the LibriSpeech dataset \cite{panayotov2015librispeech}, following \cite{zhang2023speechtokenizer}. For acoustic training, we adopt the DAC framework \cite{kumar2023high}, extracting random 5-second segments (vs.\ 0.38 s in DAC) from each utterance, zero-padding shorter ones for uniform batch sizes. This extended duration captures richer context and improves semantic representation. Unless stated otherwise, we use nine codebooks: DAC employs nine RVQ codebooks, while HASRD combines one semantic $k$-means codebook with eight acoustic RVQ codebooks. 
The $k$-means model is trained offline on SSL latent representations using 10\% of the train-clean-100 subset.
For the ablation study in Section~\ref{sec:acoustic}, models are trained on the 100h train-clean-100 subset for 30K iterations; all other experiments use the full 960h LibriSpeech dataset with 100K iterations.
For SSL training, data is prepared using Lhotse \cite{zelasko2021lhotse}, enabling dynamic batching by total batch duration. Audio is converted to 80-dimensional mel-spectrograms via TorchAudio's Kaldi filterbank\footnote{\url{https://pytorch.org/audio/main/generated/torchaudio.compliance.kaldi.fbank.html}}, using a 25~ms window with 8~ms frame shift. \\ 
\textbf{Encoder:}  
Our encoder consists of a CNN module described in Section~\ref{sec:ssl} followed by Conformer layers for enhanced semantic modeling. The CNN module consists of two 1D-CNN blocks, each with a kernel size of 4, a stride of 2, and a latent dimension of 1024. This results in an overall downsampling factor of \(4 \times 160 = 512\), where the factor of 160 originates from the 8~ms frame shift during feature extraction. This matches the downsampling of the DAC encoder. Each CNN block contains three residual units, each with a 1D-CNN and dilation rates of \(\{1, 3, 9\}\). The Conformer module follows \cite{whetten2024open}, with modifications to reduce parameters due to GPU memory constraints. Specifically, we use 12 layers instead of 24 as in \cite{chiu2022self}, increase the number of attention heads from 8 to 12, and reduce both the feedforward dimension from 2048 to 1024 and the attention dimension from 768 to 512. The encoder is pre-trained on the full LibriSpeech dataset using the MLM objective. In our proposed approach, we found that layer $n_s=8$ yields the best ASR performance, which we 
use as the semantic representation $\mathbf{H}^s$.
\\
\textbf{Decoder:} The decoder follows \cite{kumar2023high}, consisting of 4 1D-CNN upsampling layers with strides [8, 8, 4, 2] and a latent dimension of 1536. We also adopt the same multi-period and multi-scale STFT discriminators from \cite{kumar2023high}. Acoustic training uses a batch size of 8, while MLM training uses a batch duration of 600 seconds per GPU across 8 A40 GPUs. SSL training is optimized with AdamW, a learning rate of 0.0008, a Noam scheduler with 25K warmup steps, and gradient accumulation of 2. Reconstruction training is optimized with AdamW, a learning rate of $1 \times 10^{-4}$ and exponential learning rate schedule. \\
\textbf{Evaluation:} 
We use ASR as the downstream task to evaluate the quality of the semantic tokens learned by our model. We use a 2-layer BLSTM with a hidden dimensionality of 1024 as our ASR encoder. It is trained using tuples $(x, y)$, where $x$ is the learned continuous or discrete representation outputted by the pre-trained models, and $y$ is the character transcript. The model is trained using CTC \cite{graves2006connectionist} on the train-clean-100 subset of LibriSpeech. We train the model for 30K iterations using the Adam optimizer with a learning rate of 1e-4 and a batch size of 32. We report the word error rate (WER) and character error rate (CER) on the LibriSpeech test-clean subset.

For reconstruction evaluation, we compute the loss using log-mel spectrogram distances across window lengths \([32, 64, 128, 256, 512, 1024, 2048]\) with corresponding mel bins \([5, 10, 20, 40, 80, 160, 320]\). Additionally, we utilize ViSQOL, a perceptually-motivated model for measuring audio quality, to provide a more comprehensive assessment of reconstruction performance. All hyperparameters and analysis choices were selected on the dev-clean subset, while final results are reported on the test-clean subset of LibriSpeech.

\section{Results}
\subsection{Improved Computational Efficiency of DAC}\label{sec:acoustic}
We conducted ablation studies on 100 hours of LibriSpeech to evaluate the impact of our enhanced CNN encoder on reconstruction quality and efficiency using fbank inputs. Unlike previous DAC models trained on 0.38-second audio segments, we found that increasing the window size improves contextual modeling and enhances reconstruction performance, as shown in Table~\ref{tab:efficient}. Furthermore, DAC+fbank reduces GPU memory usage by 25\%, enabling a $3\times$ increase in batch size compared to raw audio. Inference speed also improves by $2\times$ on CPU, while the total model parameters decrease by 11.2\%, as only half of the encoder parameters are used. Furthermore, introducing depthwise separable convolutions exclusively in the residual CNN layers enhances inference speed by $3\times$ compared to the baseline and further reduces parameters by an additional 29\% (compared to DAC+fbank), with only a minor degradation in reconstruction quality relative to the DAC baseline. Finally, we compare the reconstruction quality of our proposed representation factorization approach to SpeechTokenizer in Table~\ref{tab:efficient2}. By increasing the number of RVQ codebooks from 8 to 9, we raise the bitrate to 3.1 kbps. Despite operating at half the bitrate of SpeechTokenizer, our method consistently achieves superior performance, reducing the mel distance from 0.76 to 0.64, STFT loss from 1.58 to 1.48, and improving the ViSQOL score from 4.26 to 4.50. Additionally, we demonstrate the generalizability of our approach to other SSL models, such as HuBERT, achieving reconstruction quality on par with SpeechTokenizer using a configuration of 9 codebooks (8 RVQ + 1 k-means).
 \begin{table}[tb!]
    \centering
     \caption{Comparative Analysis of Reconstruction Performance}
     \vspace{-9pt}
     \include{tables/efficient2}
 \vspace{-0.7cm}
    \label{tab:efficient2}
\end{table}
\subsection{SSL Pre-Training}\label{sec:semantic}
\begin{table*}[ht]
    \centering
    \begin{minipage}[t]{0.7\textwidth}
        \centering
        \caption{Comparative analysis of WER/CER results. For HASRD, only the encoder is used for ASR, but total parameters are reported for system-level comparison.}
     \vspace{-8pt}
    \include{tables/semantics}
        \label{tab:semantics} 
    \end{minipage}%
    \hspace{0.02\textwidth} 
    \begin{minipage}[t]{0.24\textwidth}
        \centering
     \caption{Speaker similarity and ASR performance using semantic and acoustic tokens}
     \vspace{-8pt}
    \include{tables/spk_sim}
    \label{tab:spk}
    \end{minipage}
    \vspace{-0.6cm}
\end{table*}

In this section, we evaluate the impact of enhancing the proposed CNN encoder from Section~\ref{sec:acoustic} with attention mechanisms and self-supervised pre-training, as shown in Table~\ref{tab:semantics}. First, we highlight the limitations of state-of-the-art high-fidelity codecs like DAC by assessing their representations on the ASR task. It can be seen that fbank features achieve an 18.3\% relative improvement in WER over DAC’s latent encoder representation, and DAC’s discretized representation proves ineffective, yielding a WER of 92\%. Following the methodology in \cite{chiu2022self}, we first reproduce the BestRQ model and introduce key improvements. Specifically, we replace the 2D-CNN pre-encoder with our optimized 1D-CNN pre-encoder from Section~\ref{sec:acoustic}, with modifications on the encoder from Section~\ref{sec:setup}. These modifications result in a 23.7\% relative WER improvement while reducing the model size to 62M parameters. Additionally, we find that switching from global normalization to sentence normalization reduces feature variance in the latent space from an average of 6 to 1, further improving WER by 4.5\% for latent representations and 14.4\% for quantized representations with k-means. Increasing the attention dimension from 516 to 768 in HASARD (BestRQ$+$) yields additional relative WER improvements of 5\% for latent representations and 25\% for quantized representations compared to BestRQ$+$, surpassing SpeechTokenizer. Notably, SpeechTokenizer suffers significant WER degradation compared to its teacher model, HuBERT base, due to the distillation process. In contrast, our approach with Hubert, HASRD (HuBERT), effectively retains the performance of pre-trained SSL models.



\subsection{Information Disentanglement}\label{sec:Disentanglement}
\begin{figure}[tb!]
\centering
\includegraphics[width=1\columnwidth]{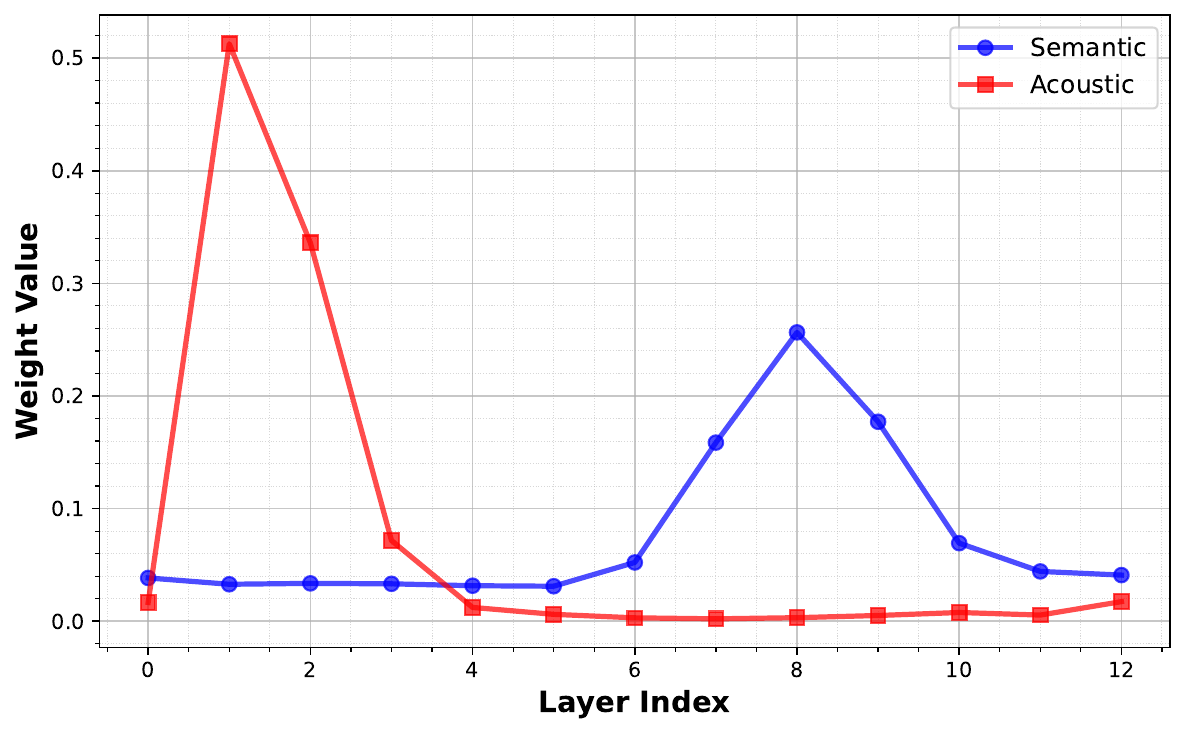}
\vspace{-.7cm}
\caption{Weights of latent representations of SSL in acoustic (reconstruction) and semantic (ASR) tasks.}
\label{fig:weighted} 
 \vspace{-0.7cm}
\end{figure}

In this section, we examine the disentanglement of acoustic and semantic information in HASRD. A key advantage of our method is its interpretability, as the acoustic representation is formed as a convex combination of SSL latent representations, with learned weights as described in Section~\ref{sec:factorization}. Figure~\ref{fig:weighted} visualizes the weights $\alpha$ learned from the speech reconstruction task and a separate set optimized on the ASR task (Section~\ref{sec:setup} \textbf{Evaluation}), showing the contributions of SSL latent representations to acoustic (reconstruction) and semantic (ASR) objectives. The representations begin with the CNN encoder, corresponding to index $0$.
The plot reveals that for speech reconstruction, the first three attention layers contribute the most, with layer indexed 1 receiving the highest weight. Interestingly, the CNN representation has a relatively low weight. We hypothesize that the first attention layer effectively integrates the information encoded by the CNN while capturing contextual dependencies and long-term relationships, reducing the direct contribution of the CNN representation. In contrast, for ASR, the highest weights are observed in layers 7–9, with layer 8 being the most influential, aligning with our ASR experiments. This analysis empirically demonstrates the inherent conflict between acoustic and semantic objectives, where improving one often degrades the other. Additionally, upon listening to the reconstructed speech using only semantic tokens (from the k-means codebook), we noticed that the speech remains intelligible but exhibits a robotic quality, making speaker identification challenging. Conversely, when reconstructing speech using only acoustic tokens (from the remaining RVQ codebooks), speaker identity is well-preserved, but intelligibility is lost. We hypothesize that our disentanglement approach effectively separates semantic and speaker-specific information, with the first codebook capturing linguistic content while the acoustic codebooks encode speaker characteristics. To validate this, we conduct a speaker similarity analysis using the ECAPA-TDNN speaker verification model \cite{desplanques2020ecapa}, measuring the cosine similarity between reference and reconstructed speech representations. Additionally, we evaluate ASR performance using both acoustic and semantic tokens to assess their respective contributions, as shown in Table \ref{tab:spk}. 
The results show that acoustic tokens achieve a high speaker similarity score (0.67) but poor ASR performance (72\% WER), whereas semantic tokens yield lower speaker similarity (0.15) but better ASR performance (21\% WER). This confirms that our approach effectively disentangles semantic content (first codebook) from speaker information (acoustic codebooks).

\section{Conclusion}
In this paper, we proposed a novel framework for learning hierarchical disentangled acoustic and semantic representations. Our approach effectively preserves semantic performance for ASR while achieving reconstruction quality comparable to state-of-the-art neural audio codecs like DAC. Additionally, we introduced an efficient CNN pre-encoder that reduces computational complexity without sacrificing quality, and presented an improved BestRQ design with better ASR performance and enhanced quantized representations. Our analysis provided insights into the inherent
conflict between acoustic and semantic objectives, demonstrating that our method successfully disentangles speaker-specific
details from linguistic content. Finally, speaker similarity and ASR evaluations further validated the effectiveness of our disentanglement strategy.

\section{References}
\label{sec:refs}
\printbibliography
\end{document}

%% file: sourcemap_definitions.tex
\DeclareSourcemap{
  \maps[datatype=bibtex, overwrite=true]{
    \map{
      \step[fieldsource=booktitle, match=\regexp{.*Interspeech.*}, replace={Proc. Interspeech}]
      \step[fieldsource=journal, match=\regexp{.*INTERSPEECH.*}, replace={Proc. Interspeech}]
      \step[fieldsource=booktitle, match=\regexp{.*ICASSP.*}, replace={Proc. ICASSP}]
      \step[fieldsource=booktitle, match=\regexp{.*icassp_inpress.*}, replace={Proc. ICASSP (in press)}]
      \step[fieldsource=booktitle, match=\regexp{.*Acoustics,.*Speech.*and.*Signal.*Processing.*}, replace={Proc. ICASSP}]
      \step[fieldsource=booktitle, match=\regexp{.*International.*Conference.*on.*Learning.*Representations.*}, replace={Proc. ICLR}]
      \step[fieldsource=booktitle, match=\regexp{.*International.*Conference.*on.*Computational.*Linguistics.*}, replace={Proc. COLING}]
      \step[fieldsource=booktitle, match=\regexp{.*SIGdial.*Meeting.*on.*Discourse.*and.*Dialogue.*}, replace={Proc. SIGDIAL}]
      \step[fieldsource=booktitle, match=\regexp{.*International.*Conference.*on.*Machine.*Learning.*}, replace={Proc. ICML}]
      \step[fieldsource=booktitle, match=\regexp{.*North.*American.*Chapter.*of.*the.*Association.*for.*Computational.*Linguistics:.*Human.*Language.*Technologies.*}, replace={Proc. NAACL}]
      \step[fieldsource=booktitle, match=\regexp{.*Empirical.*Methods.*in.*Natural.*Language.*Processing.*}, replace={Proc. EMNLP}]
      \step[fieldsource=booktitle, match=\regexp{.*Association.*for.*Computational.*Linguistics.*}, replace={Proc. ACL}]
      \step[fieldsource=booktitle, match=\regexp{.*Automatic.*Speech.*Recognition.*and.*Understanding.*}, replace={Proc. ASRU}]
      \step[fieldsource=booktitle, match=\regexp{.*Spoken.*Language.*Technology.*}, replace={Proc. SLT}]
      \step[fieldsource=booktitle, match=\regexp{.*Speech.*Synthesis.*Workshop.*}, replace={Proc. SSW}]
      \step[fieldsource=booktitle, match=\regexp{.*workshop.*on.*speech.*synthesis.*}, replace={Proc. SSW}]
      \step[fieldsource=booktitle, match=\regexp{.*Advances.*in.*neural.*information.*processing.*}, replace={Proc. NeurIPS}]
      \step[fieldsource=booktitle, match=\regexp{.*Advances.*in.*Neural.*Information.*Processing.*}, replace={Proc. NeurIPS}]
      \step[fieldsource=booktitle, match=\regexp{.*Workshop.*on.*Applications.*of.*Signal.*Processing.*to.*Audio.*and.*Acoustics.*}, replace={Proc. WASPAA}]
      \step[fieldsource=publisher, match=\regexp{.+}, replace={{}}]
      \step[fieldsource=month, match=\regexp{.+}, replace={{}}]
      \step[fieldsource=location, match=\regexp{.+}, replace={{}}]
      \step[fieldsource=address, match=\regexp{.+}, replace={{}}]
      \step[fieldsource=organization, match=\regexp{.+}, replace={{}}]
    }
  }
}

%% file: tables/efficient2.tex
\setlength{\tabcolsep}{2pt}
\resizebox{0.47\textwidth}{!}{
\begin{tabular}{l c c c c c}
\toprule
\textbf{Model (Enc.)}  & \textbf{Bitrate} & \textbf{Params} & \textbf{Mel $\downarrow$} & \textbf{STFT $\downarrow$} & \textbf{ViSQOL $\uparrow$} \\
\midrule

SpeechTokenizer        & 6.0 kbps   & 103.7M   & 0.76 & 1.58 & 4.26 \\
HASRD (BestRQ$+$)            & 3.1 kbps & \phantom{1}99.5M    & \textbf{0.64} & \textbf{1.48} & \textbf{4.50} \\
HASRD (HuBERT)       & 4.5 kbps    & 131.7M    & 0.76     & 1.57   & 4.30   \\
\bottomrule
\end{tabular}
}


%% file: tables/semantics.tex
 


     \sisetup{
    detect-weight, 
    mode=text, 
    tight-spacing=true,
    round-mode=places,
    round-precision=1,
    table-format=2.1,
    table-number-alignment=center
    }
\resizebox{1\textwidth}{!}
{
\begin{tabular}{l c c c *{4}{S}}
\toprule
&&&& \multicolumn{2}{c}{\textbf{Latent}} & \multicolumn{2}{c}{\textbf{Quantized}} \\
\cmidrule(lr){5-6} \cmidrule(lr){7-8}
\multirow{1}{*}{\textbf{Model}} & \multirow{1}{*}{\textbf{Params}} & \multirow{1}{*}{\textbf{Norm}} & \multirow{1}{*}{\textbf{CNN Module}} 
& \textbf{CER} & \textbf{WER} & \textbf{CER} & \textbf{WER} \\
\midrule
Fbank & - & - & - & 11.8 & 35.8 & {-} & {-} \\
DAC & 76.6M & - & 1D-CNN & 14.6 & 43.8 & 52.6 & 92.6 \\
\midrule
BestRQ (baseline) \cite{whetten2024open} & 106.0M & global & 2D-CNN & 4.9 & 17.3 & 12.8 & 38.3 \\
BestRQ$+$ & 62.0M & global & 1D-CNN+SepConv & 3.7 & 13.2 & 12.6 & 37.4 \\
BestRQ$+$ & 62.0M & sentence & 1D-CNN+SepConv & 3.6 & 12.6 & 9.3 & 28  \\
\midrule
SpeechTokenizer & 103.7M & - & - & 6.5 & 21.5 & 7.1 & 23.0 \\
HASRD (BestRQ$+$)& 141.6M & sentence &1D-CNN+SepConv & 3.4 & 12.0 & 6.4 & 21.0 \\
HASRD (HuBERT) & 131.7M & - &- & 2.1 & 7.4 & 3.3 & 11.3 \\
\bottomrule
\end{tabular}
}

%% file: tables/spk_sim.tex
\resizebox{1\textwidth}{!}
{
\begin{tabular}{l c c c }
\toprule
\textbf{Tokens}  & \textbf{SIM $\uparrow$} & \textbf{WER $\downarrow$}\\

\midrule

Combined         & 0.92   & 24.2 \\
Acoustic       & 0.67 & 70.2  \\
Semantic     &  0.15   &  21.0    \\
\bottomrule
\end{tabular}
}